\documentclass[twocolumn,showpacs,preprintnumbers,amsmath]{revtex4}

\usepackage{graphicx}
\usepackage{dcolumn}
\usepackage{bm}
\begin{document}

\title{Quantum phase transition of the one-dimensional transverse field compass model}
\author{Ke-Wei Sun$^{1,3}$ and Qing-Hu Chen$^{2,1,*}$}
\address{$^{1}$ Department of Physics, Zhejiang University, Hangzhou 310027,
P. R. China \\
$^{2}$ Center for Statistical and Theoretical Condensed Matter
Physics, Zhejiang Normal University, Jinhua 321004, P. R. China \\
$^{3}$ Institute of Materials Physics, Hangzhou Dianzi University,
Hangzhou 310018, China}
\date{\today}
\begin{abstract}
The quantum phase transition (QPT) of the one-dimensional (1D)
quantum compass model in a transverse magnetic field is studied in
this paper. An exact solution is obtained by using an extended
Jordan and Wigner transformation to the pseudo-spin operators. The
fidelity susceptibility, the concurrence, the block-block
entanglement entropy, and the pseudo-spin correlation functions are
calculated with antiperiodic boundary conditions. The QPT driven by
the transverse field only emerges at zero field  and is of the
second-order. Several critical exponents obtained by finite size
scaling analysis are  the same as those in the 1D transverse field
Ising model, suggesting the same universality class.  A logarithmic
divergence of  the entanglement entropy of a block  at the quantum
critical point is also observed.  From the calculated coefficient
connected to the central charge of the conformal field theory, it is
suggested that the block entanglement depends crucially on the
detailed topological structure of a system.

\end{abstract}


 \maketitle

\section{introduction}

The quantum compass model has been studied extensively in recent
years due to the possible long range orbital order and the quantum
phase transitions (QPTs)\cite
{Dorier,HDChen,Doucot,Mishra,Brzezicki,Ors,Sun,Eriksson}. First,
the model could be used to describe the Mott insulators with orbit
degeneracies. It depends on the lattice geometry, and belongs to
the low energy Hamiltonian originated from the magnetic
interactions in Mott-Hubbard systems with the strong spin-orbit
coupling\cite{You,Jackeli}. For simplicity, the 1D quantum compass
model is regarded as the coupling along one of bonds which shows
an Ising type, but different spin components are active along
other bond directions. It is exactly the same as the 1D reduced
Kitave model\cite {Feng,Shi,Bombin,Julien}. The symmetry of the
pseudo-spin Hamiltonian is much lower than SU(2). It is shown in
the numerical results that the eigenstates are at least twofold
degenerate or highly degenerate\cite {Dorier,Brzezicki}. Recently,
because of degeneracy in the ground-state (GS), the protected
qubit can be implemented, a scalable and error-free scheme of the
quantum computation can be designed in this simple
model\cite{Doucot}.

To shed some insights into this model, a exact solution is clearly
desirable. By applying Jordan and Wigner transformation to the
pseudo-spin operators, Brzezicki \emph{et al.} were able to map it
into a spinless fermion model and determine the spectrum
exactly\cite{Brzezicki}.  More recently, the exact solution of 1D
period-two compass model has also been obtained by the present
authors and a collaborator with a slightly different
method\cite{Sun}. In order to be useful for quantum information,
the GS must be protected from local perturbations, but the
spectrum for $h=0$ is gapless in the thermodynamic limit. The
extension to finite fields is a crucial step in the search for
systems supporting naturally robust quantum
information\cite{Scarola}. To the best of our knowledge, the 1D
compass model in the transverse magnetic field has not been
studied so far, which may be also of fundamental significance. The
symmetry of the system is further broken when the transverse
magnetic field is applied. The behavior of the energy gap may be
changed around the critical point in the thermodynamical limit and
the degeneracy in the GS may be lifted, therefore the nature of
the QPT may be altered in the presence of the transverse magnetic
field.  In order to address these questions, an  exact solution to
the field version is also clearly called for.

Due to the recent progress in quantum information science, some
concepts in quantum information theory, such as the fidelity, the
fidelity susceptibility (FS), and the quantum entanglement have
been extensively used to identify the QPTs in various many-body
systems from the perspective of the GS wave functions\cite
{Buonsante,Cozzini,Chen,Preskill,Osborne,Vidal,Korepin,Kitaev,Verstraete}.
Recently, it is proposed that the fidelity approach is a valuable
tool to investigate novel phases lacking a clear characterization
in terms of local order parameters\cite{Zhou,Abasto}. With these
effective tools and the finite-size scaling analysis of the FS,
one can identify the universality class of the QPT in various
models\cite{Gu,liu}. Quantum entanglement is one of the most
striking consequences of quantum correlation in many-body systems,
and is recognized to be resource that enables quantum computing
and communication\cite{Nielsen}. It has shown a deep relation with
the QPT\cite{Osterloh,Zhang}. The entangled degree between any two
nearest-neighbor particles keeps the same for the translational
symmetry, and its derivative may play the role of an order
parameter to characterize QPT at the critical point. In the
context of QPTs, the quantum entanglement has been the subject of
considerable interests in the various
models\cite{Emary,Liberti,chenqh,Osborne,Vidal}.

In this paper, we study the 1D compass model in a transverse
magnetic field  for the first time. The exact solutions are
obtained by using the method of mapping into a case with plural
spin sites\cite {Sasaki}. The GS fidelity, the FS, the concurrence
and the block-block entanglement entropy are calculated. The
behaviors of the spin correlations function are also given. The
paper is organized as follows: In Section \textbf{II}, we describe
the model and the scheme to obtain  the exact solution in detail.
The calculations of the fidelity, the concurrence and the
block-block entanglement entropy are presented in Section
\textbf{III}, where the scaling analysis is also performed. The
correlation functions are analyzed in Section \textbf{IV}. The
conclusion is given in the last section.

\section{MODEL HAMILTONIAN AND EXACT SOLUTION}

The 1D XX-YY model in a transverse magnetic field can be regarded
as the structure of two pseudo-spin sites inside a unit cell. The
Hamiltonian is given by\cite{Brzezicki}
\begin{eqnarray}
H&=&-J\sum_{n=1}^{N^{\prime }}\sigma _{2,n}^x\sigma
_{1,n+1}^x-J(1-\beta)\sum_{n=1}^{N^{\prime }}\sigma
_{1,n}^x\sigma _{2,n}^x\nonumber \\
&&-J\beta\sum_{n=1}^{N^{\prime }}\sigma _{1,n}^y\sigma
_{2,n}^y-\frac h2\sum_{n=1}^{N^{\prime
}}(\sigma_{1,n}^z+\sigma_{2,n}^z),
\end{eqnarray}
where $N=2N^{\prime }$ is the total number of the sites. Fig. 1
shows the structure of interactions in Eq. (1).

\begin{figure}[tbp]
\includegraphics[scale=1.0]{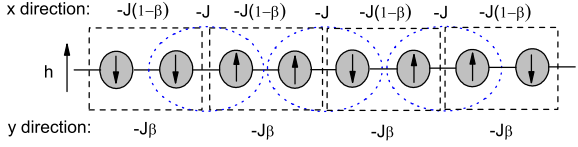}
\caption{(Color online) The odd \{1,n;2,n\} and the even
\{2,n;1,n+1\} bonds are denoted in the regions of  black dashed
rectangle and blue dotted ellipse, respectively.} \label{spin
structure}
\end{figure}

For $\beta=1$, it becomes the 1D compass model in a transverse
magnetic field
\begin{eqnarray}
H&=&-J\sum_{n=1}^{N^{\prime }}\sigma _{2,n}^x\sigma
_{1,n+1}^x-J\alpha\sum_{n=1}^{N^{\prime }}\sigma _{1,n}^y\sigma _{2,n}^y
\nonumber \\
&&-\frac h2\sum_{n=1}^{N^{\prime }}(\sigma_{1,n}^z+\sigma_{2,n}^z),
\end{eqnarray}
where $J$ denotes the strength of  the nearest-neighbor interaction,
$\alpha $ is the coupling parameter, $\sigma _{s,n}^{x(y,z)}$ are
the Pauli matrix on cell $n$ with site $s=1, 2$, $N=2N^{\prime }$ is
the total number of the sites, and $h$ is the applied magnetic field
in the $z$ direction. For convenience, the number of pseudospins
$N^{\prime}$ is chosen to be even, and a periodic boundary
conditions (PBC) for pseudospins is employed, i.e.
$\sigma_{1,N^{\prime}+1}=\sigma_{1,1}$. Note that the 1D compass
model without the magnetic field is just a special case of the
alternating XY model\cite{Perk}.

In order to diagonalize the Hamiltonian (2), we use the extension of
the Jordan and Wigner transformation for the case with plural spin
sites\cite{Sasaki}. An up-spin state is transformed to a one-fermion
state, and a down-spin state to a zero-fermion state. The explicit
mapping between spin operators and fermionic operators are given by
\begin{eqnarray}
\sigma _{2,n}^x\sigma
_{1,n+1}^x=(a_{2,n}^{\dagger}-a_{2,n})(a_{1,n+1}^{\dagger}+a_{1,n+1})\nonumber\\
\sigma _{1,n}^y\sigma
_{2,n}^y=-(a_{1,n}^{\dagger}+a_{1,n})(a_{2,n}^{\dagger}-a_{2,n})\nonumber\\
\sigma_{s,n}^z=2a_{s,n}^{\dagger}a_{s,n}-1.
\end{eqnarray}
Here we denote the fermion creation operator with site number $s$
and cell number $n$ by $a_{s,n}^{\dagger}$. Then the Hamiltonian (2)
is transformed into the following form
\begin{eqnarray}
H&=&-J\sum_{n=1}^{N^{\prime
}}(a_{2,n}^{\dagger}a_{1,n+1}^{\dagger}+a_{2,n}^{
\dagger}a_{1,n+1}-a_{2,n}a_{1,n+1}^{\dagger}  \nonumber \\
&&-a_{2,n}a_{1,n+1})-J\alpha\sum_{n=1}^{N^{\prime
}}(-a_{1,n}^{\dagger}a_{2,n}^{\dagger}+a_{1,n}^{\dagger}a_{2,n}  \nonumber \\
&&-a_{1,n}a_{2,n}^{\dagger}+a_{1,n}a_{2,n})-h\sum_{n=1}^{N^{\prime
}}(a_{1,n}^{\dagger}a_{1,n}  \nonumber \\
&&+a_{2,n}^{\dagger}a_{2,n})+hN^{\prime}.
\end{eqnarray}
The Fourier transformation of the fermion operators gives $
a_{s,n}=(1/N^{\prime})^{1/2}\sum_{p}e^{-ipn}a_{s}(p)$. For
convenience, the antiperiodic boundary condition (ABC)
$a_{1,N^{\prime}+1}=-a_{1,1}$ is employed for the  fermion
operators. After these transformations, the new Hamiltonian
$H^{\prime}$ now reads
\begin{eqnarray}
H^{\prime}&=&-J\sum_{p}[e^{-ip}a_{2}^{\dagger}(p)a
_{1}^{\dagger}(-p)+e^{-ip}a_{2}^{\dagger}(p)a _{1}(p)  \nonumber \\
&&-e^{ip}a_{2}(p)a _{1}^{\dagger}(p)-e^{ip}a_{2}(p)a _{1}(-p)]  \nonumber \\
&&-J\alpha\sum_{p}[-a_{1}^{\dagger}(p)a
_{2}^{\dagger}(-p)+a_{1}^{\dagger}(p)a _{2}(p)  \nonumber \\
&&-a_{1}(p)a _{2}^{\dagger}(p) -a_{1}(p)a _{2}(-p)]  \nonumber \\
&&-h\sum_{p}[a_{1}^{\dagger}(p)a _{1}(p)+a_{2}^{\dagger}(p)a
_{2}(p)]+hN^{\prime},
\end{eqnarray}
where $p$ is the wave number in ABC which takes such values as
$p=\pm j\pi/N^{\prime }, (j=1,3,...N^{\prime }-1)$. The operators
$a_s^{\dagger}(p)$ and $a_s(p)$ are the creation and annihilation
operators of the fermion with site numbers $s$ and wave number
$p$, which satisfy the following anticommutation relations
\begin{eqnarray}
\{a_s(p),a_t^{\dagger}(q)\}=\delta_{s,t}\delta_{p,q},  \nonumber \\
\{a_s(p),a_t(q)\}=0,\{a_s^{\dagger}(p),a_t^{\dagger}(q)\}=0.
\end{eqnarray}
Then we find that the Hamiltonian $H^{\prime}$ is the sum of the
following  independent operators $W(p^{\prime})$:
\begin{eqnarray}
W(p^{\prime})&=&-J[e^{-ip^{\prime}}a_{2}^{\dagger}(p^{\prime})a
_{1}^{\dagger}(-p^{\prime})+e^{-ip^{\prime}}a_{2}^{\dagger}(p^{\prime})a
_{1}(p^{\prime})  \nonumber \\
&&-e^{ip^{\prime}}a_{2}(p^{\prime})a
_{1}^{\dagger}(p^{\prime})-e^{ip^{\prime}}a_{2}(p^{\prime})a
_{1}(-p^{\prime})  \nonumber \\
&&+e^{ip^{\prime}}a_{2}^{\dagger}(-p^{\prime})a
_{1}^{\dagger}(p^{\prime})+e^{ip^{\prime}}a_{2}^{\dagger}(-p^{\prime})a
_{1}(-p^{\prime})  \nonumber \\
&&-e^{-ip^{\prime}}a_{2}(-p^{\prime})a
_{1}^{\dagger}(-p^{\prime})-e^{-ip^{\prime}}a_{2}(-p^{\prime})a
_{1}(p^{\prime})]  \nonumber \\
&&-J\alpha[-a_{1}^{\dagger}(p^{\prime})a
_{2}^{\dagger}(-p^{\prime})+a_{1}^{\dagger}(p^{\prime})a _{2}(p^{\prime})
\nonumber \\
&&-a_{1}(p^{\prime})a _{2}^{\dagger}(p^{\prime}) -a_{1}(p^{\prime})a
_{2}(-p^{\prime})  \nonumber \\
&&-a_{1}^{\dagger}(-p^{\prime})a
_{2}^{\dagger}(p^{\prime})+a_{1}^{\dagger}(-p^{\prime})a _{2}(-p^{\prime})
\nonumber \\
&&-a_{1}(-p^{\prime})a _{2}^{\dagger}(-p^{\prime}) -a_{1}(-p^{\prime})a
_{2}(p^{\prime})]  \nonumber \\
&&-h[a_{1}^{\dagger}(p^{\prime})a
_{1}(p^{\prime})+a_{2}^{\dagger}(p^{\prime})a _{2}(p^{\prime})  \nonumber \\
&&+a_{1}^{\dagger}(-p^{\prime})a
_{1}(-p^{\prime})+a_{2}^{\dagger}(-p^{\prime})a _{2}(-p^{\prime})],
\end{eqnarray}
where $p^{\prime}=j\pi/N^{ \prime }, (j=1,3,...N^{\prime }-1)$. Note
that $[W(p^{\prime}),W(q^{\prime})]=0$, so we can  solve the
Hamiltonian (5) in the space of $p^{\prime}$.

The parity in the Hilbert space of $W(p^{\prime})$  is conserved,
therefore subspace with the even parity  can be easily constructed
in terms of the following 8 basis vectors $|0\rangle ,a_1^{\dagger
}(p^{\prime })a_1^{\dagger }(-p^{\prime })|0\rangle ,a_1^{\dagger
}(p^{\prime })a_2^{\dagger }(-p^{\prime })|0\rangle ,a_2^{\dagger
}(p^{\prime })a_1^{\dagger }(-p^{\prime })|0\rangle
,\nonumber\newline a_2^{\dagger }(p^{\prime })a_1^{\dagger
}(-p^{\prime })|0\rangle ,a_1^{\dagger }(p^{\prime })a_2^{\dagger
}(p^{\prime })|0\rangle ,a_1^{\dagger }(-p^{\prime })a_2^{\dagger
}(-p^{\prime })|0\rangle ,$ and$ \nonumber\newline $ $a_2^{\dagger
}(p^{\prime })a_1^{\dagger }(p^{\prime })a_2^{\dagger }(-p^{\prime
})a_1^{\dagger }(-p^{\prime })|0\rangle $. While  the subspace with
the odd parity  is obtained by combining the following 8 basis
vectors $a_1^{\dagger }(p^{\prime })|0\rangle, a_2^{\dagger
}(p^{\prime })|0\rangle, a_1^{\dagger }(-p^{\prime })|0\rangle,
a_2^{\dagger }(-p^{\prime })|0\rangle, a_1^{\dagger }(-p^{\prime
})a_2^{\dagger }(p^{\prime })\nonumber\newline a_1^{\dagger
}(p^{\prime })|0\rangle, a_2^{\dagger }(-p^{\prime })a_2^{\dagger
}(p^{\prime })a_1^{\dagger }(p^{\prime })|0\rangle, a_1^{\dagger
}(p^{\prime })a_2^{\dagger }(-p^{\prime })a_1^{\dagger }(-p^{\prime
})|0\rangle,$ and $a_2^{\dagger }(p^{\prime })a_2^{\dagger
}(-p^{\prime })a_1^{\dagger }(-p^{\prime })|0\rangle$. The parity of
subspaces determines the boundary conditions. Indeed, the Bogoliubov
vacuum has even (odd) numbers of $a$ quasiparticles for ABC
(PBC)\cite{Brzezicki}. For ABC, nonzero elements of the $8\times8$
Hermit matrix (even parity) for the reduced Hamiltonian
$W(p^{\prime})$ are
\begin{eqnarray}
W_{i,i}=-2h\quad\mbox for 1\le i\le 7, W_{8,8}=-4h,\nonumber\\
W_{2,3}=W_{4,5}=-Je^{-ip^{\prime}}-J\alpha,\nonumber\\
W_{1,3}=W_{4,8}=Je^{-ip^{\prime}}+J\alpha,\nonumber\\
W_{2,4}=W_{3,5}=W_{1,4}=W_{3,8}=-Je^{ip^{\prime}}-J\alpha.
\end{eqnarray}
The eigenvalues for this matrix are then easily derived
\begin{eqnarray}
\lambda ^{(1,2)}(p^{\prime }) &=&-2h\pm 2J[(h/J)^2+1+2\alpha \cos (p^{\prime
})+\alpha ^2]^{1/2}  \nonumber \\
\lambda ^{(3,4)}(p^{\prime }) &=&-2h\pm 2J[1+2\alpha \cos (p^{\prime
})+\alpha ^2]^{1/2}  \nonumber \\
\lambda ^{(5-8)}(p^{\prime }) &=&-2h.
\end{eqnarray}
The spectral functions $\epsilon (p^{\prime })$ are readily
obtained
\begin{eqnarray}
\epsilon ^{(1,2)}(p^{\prime }) &=&\pm2J[(h/J)^2+1+2\alpha \cos
(p^{\prime
})+\alpha ^2]^{1/2}  \nonumber \\
\epsilon ^{(3,4)}(p^{\prime }) &=&\pm2J[1+2\alpha \cos (p^{\prime
})+\alpha ^2]^{1/2}.
\end{eqnarray}
Actually, the Hamiltonian (5) can be decomposed as
\begin{eqnarray}
\mathcal{H^{\prime}}=\sum_{p^{\prime
}}\oplus\mathcal{H}_{p^{\prime }}^{(s^{\prime})},
\end{eqnarray}
where $\mathcal{H}_{p^{\prime }}^{(s^{\prime })}\equiv \epsilon
^{(s^{\prime })}(p^{\prime })\eta _{p^{\prime }}^{\dagger (s^{\prime
})}\eta _{p^{\prime }}^{(s^{\prime })}$ ($s^{\prime }=1,2,3,4$) with
$\eta _{p^{\prime }}^{(s^{\prime })}$ the operator of fermionic
quasiparticles , and the corresponding eigenvectors are $\psi
^{(s^{\prime })}(p^{\prime })=\eta _{p^{\prime }}^{\dagger
(s^{\prime })}|0\rangle $.  Then the GS energy and wave function are
given by
\begin{equation}
E_G=-\sum_{p^{\prime }}2J[(h/J)^2+1+2\alpha \cos (p^{\prime
})+\alpha ^2]^{1/2},
\end{equation}
\begin{equation}
|\psi _0\rangle =\prod_{p^{\prime }}\psi ^{(1)}(p^{\prime }).
\end{equation}
It should be stressed here that Eqs. (12) and (13) are valid for any
value of $h$. Note also that the exact spectrum is the same as that
obtained by Brzezicki \emph{et al}. for $h=0$ using a different
method\cite{Brzezicki}.

The energy gap can be readily obtained as
$\Delta=2J[(h/J)^2+1\pm2\alpha+\alpha^2]^{1/2}-2J|\alpha\pm1|$. It
does  not disappear in the presence of the transverse field even
in the thermodynamic limit.  The QPT driven by the transverse
field will occur at ($\alpha=\pm1$,$h=0$), which shows the
second-order nature, similar to the QPT driven by interaction
parameters\cite{Eriksson}.

For completeness, we will also briefly discuss the spectra based
on PBC. For PBC,  we need to solve the Hamiltonian (7) in the odd
numbers of $a$ quasiparticles subspace. The spectral functions are
given by
\begin{eqnarray}
\varepsilon ^{(1,2)}(p^{\prime })&=&-J[(h/J)^2+1+2\alpha \cos
(p^{\prime })+\alpha ^2]^{1/2}\nonumber\\
&&-J[1+2\alpha \cos(p^{\prime
})+\alpha ^2]^{1/2}  \nonumber \\
\varepsilon ^{(3,4)}(p^{\prime })&=&-J[(h/J)^2+1+2\alpha \cos
(p^{\prime })+\alpha ^2]^{1/2}\nonumber\\
&&+J[1+2\alpha \cos
(p^{\prime})+\alpha ^2]^{1/2}  \nonumber \\
\varepsilon ^{(5,6)}(p^{\prime })&=&J[(h/J)^2+1+2\alpha \cos
(p^{\prime })+\alpha ^2]^{1/2}\nonumber\\
&&-J[1+2\alpha \cos
(p^{\prime
})+\alpha ^2]^{1/2}  \nonumber \\
\varepsilon ^{(7,8)}(p^{\prime })&=&J[(h/J)^2+1+2\alpha \cos
(p^{\prime })+\alpha ^2]^{1/2}\nonumber\\
&&+J[1+2\alpha \cos (p^{\prime })+\alpha ^2]^{1/2}.
\end{eqnarray}
Note that  $p^{\prime}=0$ and $p^{\prime}=\pi$ in ABC must be
treated separately and carefully. It is helpful to write down
explicitly the spectra for $N=4$ pseudospin sites in the real space
\begin{eqnarray}
\omega^{(1,2)}&=&\pm2J[(h/J)^2+1+\alpha ^2]^{1/2}\nonumber\\
\omega^{(3,4)}&=&\pm2J[1+\alpha ^2]^{1/2}\nonumber\\
\omega^{(5-8)}&=&0\nonumber\\
\omega^{(9,10)}&=&J\alpha+J\pm J[(h/J)^2+1-2\alpha+\alpha ^2]^{1/2}\nonumber\\
\omega^{(11,12)}&=&J\alpha-J\pm J[(h/J)^2+1+2\alpha+\alpha ^2]^{1/2}\nonumber\\
\omega^{(13,14)}&=&-J\alpha-J\pm J[(h/J)^2+1-2\alpha+\alpha ^2]^{1/2}\nonumber\\
\omega^{(15,16)}&=&-J\alpha+J\pm J[(h/J)^2+1+2\alpha+\alpha ^2]^{1/2},\nonumber\\
\end{eqnarray}
which include the spectra for both ABC [$\omega^{(1-8)}] $ and PBC
[$\omega^{(9-16)}]$. The GS energy for PBC can be written as
\begin{eqnarray}
E_G&=&\sum_{p^{\prime }\not=0 or \pi}-2J[(h/J)^2+1+2\alpha \cos
(p^{\prime })+\alpha ^2]^{1/2}\nonumber\\
&&+min[\omega^{(s^{\prime})}],
\end{eqnarray}
where $s^{\prime}$=9-16, and $p^{\prime }=2j\pi/N^{\prime },
(j=1,2,...\frac{N^{\prime }}{2}-1)$.

The validity of Eq. (16) is also confirmed by comparing with the
direct numerical diagonalization of  $N=8$ pseudospin sites in real
space. In Fig. 2, we present the numerical GS energy from PBC, and
the results from Eq. (16) as well. It is clear that the present
analytical results for the GS energy are  in excellent agreement
with the numerical ones.

\begin{figure}[tbp]
\includegraphics[scale=0.5]{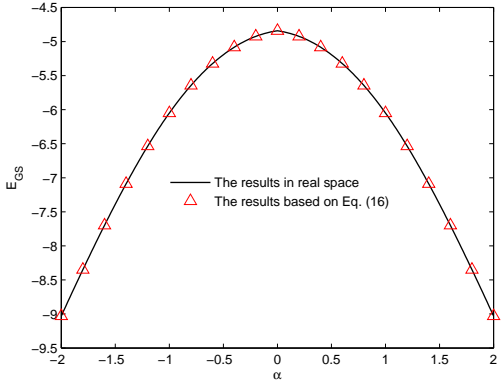}
\caption{(Color online) GS energy as the function of the parameter
$\alpha$ for $h=0.8$, $J=1.0$ and $N=8$ with PBC. } \label{EGS1}
\end{figure}

In order to show the correctness of the present method, we extend
to study the celebrated 1D Ising model in a transverse magnetic
field, which Hamiltonian reads
\begin{eqnarray}
H&=&-J\sum_{n=1}^{N^{\prime }}\sigma _{2,n}^x\sigma
_{1,n+1}^x-J\sum_{n=1}^{N^{\prime }}\sigma _{1,n}^x\sigma _{2,n}^x  \nonumber
\\
&&+\frac h2\sum_{n=1}^{N^{\prime }}(\sigma_{1,n}^z+\sigma_{2,n}^z).
\end{eqnarray}
With ABC, the exact GS energy is derived as
\begin{eqnarray}
E_G=-\sum_{p}(h^2+4J^2\pm4Jh\cos \frac p 2)^{1/2},
\end{eqnarray}
where $p=j\pi/N^{\prime }, (j=1,3,...N^{\prime }-1)$, and the
number of total sites is $N=2N^{\prime}$. For PBC, the GS energy
is written as
\begin{eqnarray}
E_G&=&-\sum_{p\not=0 or \pi}(h^2+4J^2\pm4Jh\cos \frac p
2)^{1/2}\nonumber\\
&&+min[\omega^{(s^{\prime})}],
\end{eqnarray}
where $min[\omega^{(s^{\prime})}]=-2J-(4J^2+h^2)^{1/2}$ ($2J>h>0$),
and $p=2j\pi/N^{\prime }, (j=1,2,...\frac{N^{\prime }}{2}-1)$.
Therefore, we recover the well known results obtained previously in
this model\cite{Sachdev}. It is observed that the components in the
GS energy are different for the 1D Compass and Ising models in the
transverse magnetic fields for both PBC and ABC. It should be
pointed out that although the GS in PBC and ABC are slightly
different in the finite size system, they are identical in the
thermodynamic limit and the essential features in finite-size are
also not altered  qualitatively.   Without loss of generality, we
will take ABC in the following discussion.

\section{FINITE-SIZE SCALING ANALYSIS OF FIDELITY AND ENTANGLEMENT}

The GS fidelity and entanglement emerged from quantum information
science have been used in signaling the QPTs\cite
{Buonsante,Cozzini,Chen,Abasto,Gu,Zhang,Osterloh,chenqh,liu}. We
perform finite-size scaling analysis of these two quantities to
study the criticality of the present model.  By using the exact GS
wave function obtained in Eq.(13), the GS fidelity is given by
\begin{eqnarray}
F(\alpha ,\delta\alpha )=|\langle \psi _0(\alpha )|\psi _0(\alpha
+\delta\alpha )\rangle |,
\end{eqnarray}
where $\delta\alpha $ is a small quantity ($\delta\alpha =10^{-4}$
is taken in the present calculation). Its susceptibility can be
written as
\begin{eqnarray}
\chi_F\equiv\lim_{\delta\alpha \to 0}\frac{-2lnF}{\delta\alpha^2}.
\end{eqnarray}
\begin{figure}[tbp]
\includegraphics[scale=0.55]{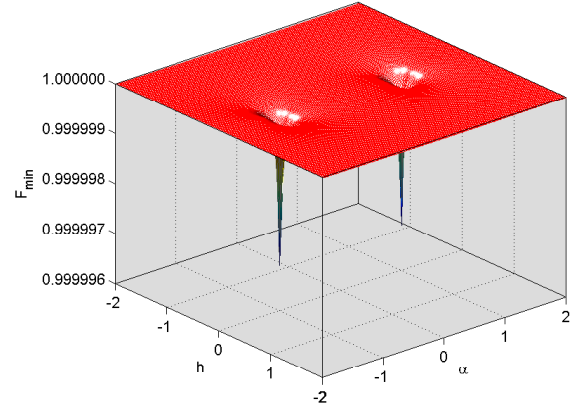}
\caption{(Color online) Fidelity $F_{min}=min[F(h,h+\delta h),
F(\alpha,\alpha+\delta \alpha)]$ in the $\alpha-h$ plane for
$N^{\prime}=100$ and $J=1.0$ with ABC. The second-order QPT points
are obviously found at ($\alpha=-1.0,h=0$) and ($\alpha=1.0,h=0$).}
\label{fidelity-h-a}
\end{figure}

\begin{figure}[tbp]
\includegraphics[scale=0.8]{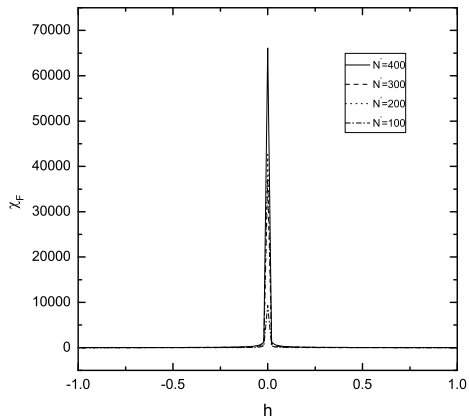}
\caption{FS versus $h$ for $J=1.0$ and $\alpha=1.0$ with ABC.}
\label{XF-h}
\end{figure}
We calculate the GS fidelity in the $(\alpha, h)$-plane, and the FS
as a function of the transverse field $h$ for $\alpha=1.0$. The
numerical results are presented in Figs . 3 and 4. The absence of
the sudden drop to zero of the fidelity excludes the level-crossing
around the critical point ($\alpha=\pm1.0$,h=0). In the
Kosterlitz-Thouless phase transition, no singularity occurs at the
critical point\cite{Chen}, so a second-order QPT is highly suggested
when driving the magnetic field, which will be confirmed in the
following finite size-scaling analysis.

Next, we illustrate the scaling behavior of average FS
$\chi_F/N^{\prime}$. The finite-size scaling ansatz for the average
FS to analyze the second-order QPT takes the form\cite{Gu,liu}
\begin{eqnarray}
\frac
{\chi_F^{max}-\chi_F}{\chi_F}=f[{N^{\prime}}^{\nu}(h-h_{max})],
\end{eqnarray}
where $\nu$ is the critical exponent of the correlation length and
f(x) is the scaling function. This function should be universal
for large $N^{\prime}$ in the second-order QPT. As exhibited  in
Fig. 4, the FS reaches a maximum point at a certain position
$h_{max}$. It can be observed in Fig. 5 that the rescaled FS for
larger system sizes tends to collapse onto one single curve if
adjusting the critical exponent $\nu=1.00\pm0.02$.  The scaled
average FS at the maximum point as a function of $N^{\prime}$ in
log-log scale are presented in the inset of Fig. 5. A power law
behavior $\chi _F^{\max } \propto {N^{\prime}}^{\mu}$ is observed
in the large $N^{\prime}$ regime and the finite-size exponent
extracted from  the curve is $\mu=2$. Both values of exponents
$\nu$ and $\mu$ in the present model are the same as those
obtained in the 1D transverse-field Ising model\cite{Chen}.

\begin{figure}[tbp]
\includegraphics[scale=0.8]{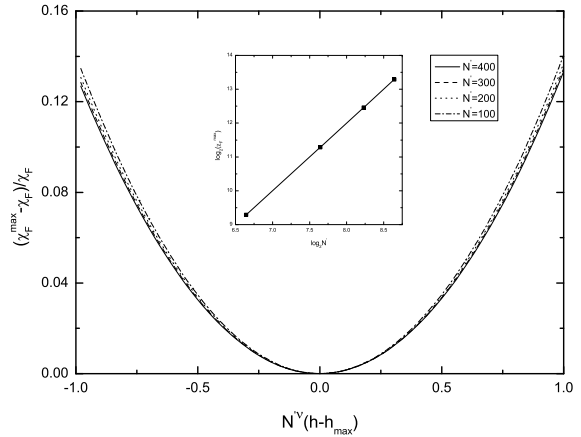}
\caption{Finite-size scaling of the average FS according to Eq.
(22) for $h_{max}=0$, $J=1.0$ and $\alpha=1.0$ for various system
sizes $N^{\prime}=100,200,300,400$. The inset exhibits the scaling
of the maximum of the average FS as the function of
$log_{2}(N^{\prime})$ at the critical point ($\alpha=1.0$,h=0).}
\label{FS-nu-scaling}
\end{figure}

We then turn to the quantum entanglement of this system. Recently,
the concept of concurrence is usually adopted as the measure of the
local entanglement in spin $-\frac 12$ systems. The definition of
concurrence is given by $
C(i,j)=max[r_1(i,j)-r_2(i,j)-r_3(i,j)-r_4(i,j),0]$, where $r_\alpha
(i,j)$ are the square roots of the eigenvalues of the product matrix
$R=\rho _{ij} \tilde{\rho _{ij}}$ in descending
order\cite{Zhang,Osterloh}. The spin
flipped matrix $\tilde{\rho _{ij}}$ is defined as $\tilde{\rho _{ij}}%
=(\sigma ^y\otimes \sigma ^y)\rho_{ij}^{*}(\sigma ^y\otimes \sigma
^y)$. The $\rho _{ij}$ is the density matrix for a pair of qubits
from a multi-qubit state, and   has the following form
\begin{eqnarray}
\rho_{ij}=\frac {1}
{4}\sum_{\alpha,\beta=0}^{3}p_{\alpha\beta}\sigma^{\alpha}_i\otimes\sigma^{
\beta}_j.
\end{eqnarray}
The coefficients are determined by the relations
\begin{eqnarray}
p_{\alpha\beta}=tr(\sigma^{\alpha}_i\sigma^{\beta}_j\rho_{ij})=\langle%
\sigma^{\alpha}_i\sigma^{\beta}_j\rangle.
\end{eqnarray}

According to the reflection symmetry and  the global phase flip
symmetry, considering the Hamiltonian being real, the only nonzero
coefficients in Eq. (24) are
$p_{00},p_{11},p_{22},p_{33},p_{03},p_{30}$. Because the density
matrix must have trace unity, so $p_{00}=1$. The numerical results
for the concurrence as a function of $h$ for various coupling
coefficient $\alpha$ are shown in Fig. 6. It is evident that the
concurrence gradually increases as enhancing $\alpha$. The minimum
of concurrence  and a cusp of the first derivative of the
concurrence occurs right at the critical point ($\alpha=1.0, h=0$).
\begin{figure}[tbp]
\includegraphics[scale=0.8]{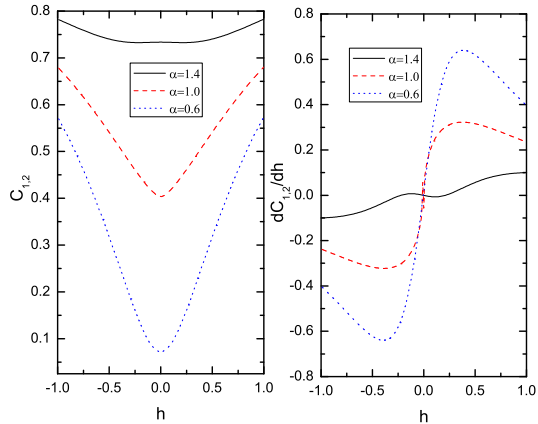}
\caption{(Color online) The concurrence $C_{1,2}$ (left) and the
derivative $\partial_{\alpha}C_{1,2}$ (right) versus $h$ for
$\alpha=0.6,1.0,1.4$, $J=1.0$ and $N^{\prime}=256$. }
\label{con-difcon}
\end{figure}

Furthermore, we calculate the block-block entanglement both near and
at the quantum critical point\cite{Vidal,Lou,Refael,Fradkin} to show
the connection with entropy of vacuum in the classical conformal
field theory in the present model. The GS in our model can be
completely characterized by the expectation values of the two-point
correlations $\langle a_m^{\dagger}a_n\rangle=f_{m,n}$, where $m$ or
$n$ is pseudospin site (e.g. $a_{s,n}\to a_{2(n-1)+s}$). Any other
expectation value can be expressed through Wick's theorem. By
eliminating the rows and columns in  matrix $F=f_{mn},
(m,n=1,2...N)$, which are corresponding to pseudospins that do not
belong to the block, the correlation matrix $F_L$ of the state
$\rho_L$ is obtained. The corresponding von Neumann entropy then
takes the form
\begin{eqnarray}
S_L=\sum_{n=1}^{L}[-(1-\lambda_n)log_2(1-\lambda_n)-\lambda_n
log_2 \lambda_n],
\end{eqnarray}
where $\lambda_n$ is the $n$th eigenvalue of the correlation matrix
$F_L$. The numerical results  for $S_L$ as a function of the block
size $L$  are presented in Fig. 7. A logarithmic divergence of $S_L$
at the quantum critical point is observed, while noncritical
entanglement is characterized by a saturation of $S_L$ for larger
$L$. The coefficient is connected to the central charge of the
classical conformal field theories,
\begin{eqnarray}
lim_{L\to\infty}S_L\sim\frac c 3 log_2 L,
\end{eqnarray}
where $c=1$ in the compass model. The value of $c$ is different from
that in 1D transverse Ising chain ($c=0.5$), but the same as in 1D
XX chain without magnetic field\cite{Vidal}.  It may follow that the
block entanglement  depends crucially on the detailed topological
structure of a system.

\begin{figure}[tbp]
\includegraphics[scale=0.8]{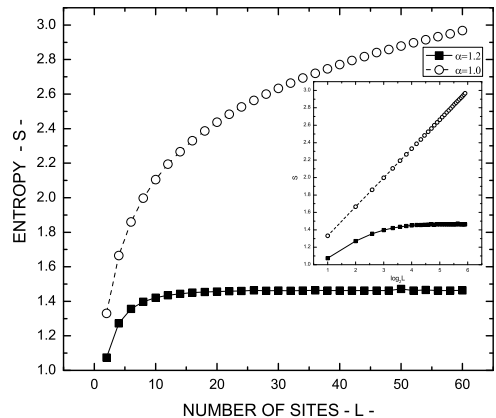}
\caption{The block-block entanglement $S_L$ versus $L$ for $J=1.0$
in the quantum critical point ($\alpha=1.0,h=0$) and the noncritical
region $\alpha=1.2$ with ABC. The inset displays a logarithmic
divergence for large $L$ at the  critical point.} \label{entropy}
\end{figure}

\section{PSEUDO-SPIN CORRELATION FUNCTIONS AND MAGNETIZATION}

To explore the essential properties of QPT, we will calculate two
GS pseudo-spin correlations  $\langle\sigma _{2,1}^x\sigma
_{1,2}^x\rangle$ and $ \langle\sigma _{1,1}^y\sigma
_{2,1}^y\rangle$. The numerical results for these two correlation
functions versus  $\alpha$  for different magnetic fields  are
presented in Fig. 8. We observe that $ \langle\sigma
_{1,1}^y\sigma _{2,1}^y\rangle$ is a odd function of $\alpha$,
while $\langle\sigma_{2,1}^x\sigma_{1,2}^x\rangle$ an even one of
$\alpha$. The crossing points of $ \langle\sigma _{2,1}^x\sigma
_{1,2}^x\rangle$ and $\langle\sigma _{1,1}^y\sigma
_{2,1}^y\rangle$ curves deviate $\alpha=1$ in the presence of
transverse field. The numerical results indicate that
$\langle\sigma_{2,1}^x\sigma_{1,2}^x\rangle$ is sensitive to the
external magnetic field in the range of $\alpha\in[-1,1]$, but
insensitive in the other regions.
\begin{figure}[tbp]
\includegraphics[scale=0.7]{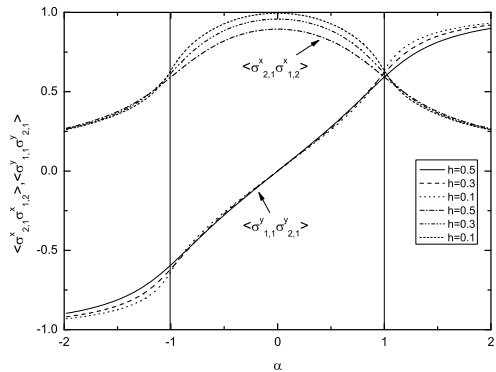}
\caption{ The correlation function with ABC. The parameters are
taken as $J=1.0$, $N^{\prime}=256$ and $h=0.1,0.3,0.5$,
respectively.} \label{correlation-function}
\end{figure}
\begin{figure}[tbp]
\includegraphics[scale=0.5]{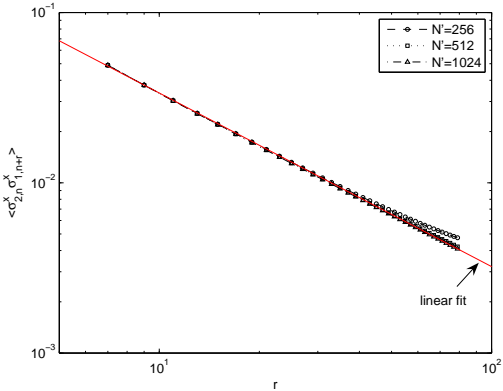}
\caption{(Color online) Distance dependence of
$\langle\sigma_{2,1}^x\sigma_{1,2+r}^x \rangle $ spin correlator at
$\beta\rightarrow 1$ for different system size. The parameters are
$J=1.0$, $h=0$ and $\alpha=1.0$.} \label{correlation1}
\end{figure}

As done in Ref. \cite{Brzezicki}, we also calculate the distance
dependence of the pseudo-spin correlator $
\langle\sigma_{2,1}^x\sigma_{1,2+r}^x\rangle$ with ABC for different
size in terms of  the Hamiltonian (1). As shown in Fig. 9 that the
correlators at $\beta\rightarrow 1$ decay in an algebraic way in
large $r$ regime, indicating a divergent correlation length when
approaching the critical points. A power law behavior
$\langle\sigma_{2,1}^x\sigma_{1,2+r}^x\rangle\propto r^{-\eta}$ is
obtained with $\eta = 1.00\pm0.03$, indicating   a second-order
 QPT.

Finally, we calculate the pseudospin magnetization
$\langle\sigma^z\rangle=\langle\sigma_{1,n}^z\rangle+\langle\sigma_{2,n}^z\rangle$
and the magnetic susceptibility $\chi=\partial
\langle\sigma^z\rangle /\partial h$. The magnetization
$\langle\sigma^z\rangle$ as a function of  the transverse field $h$
for $\alpha=1.0$ and $N^{\prime}=256$ is exhibited in Fig. 10. The
magnetic susceptibility $|\chi(h)-\chi(h_c)|$ versus $|h-h_c|$ shows
a power law behavior. The exponent $\gamma$ is estimated to be
$1.78\pm 0.05$ by the slop. It is interesting that it is very close
to the magnetic susceptibility exponent $1.75$ in 2D classical Ising
model. According to Eq. (17), we can also plot the similar scaling
curve for 1D transverse field Ising model, which is given in the
inset of Fig. 10 as well. A excellent agreement for the slop in the
critical regime is clearly shown.

\begin{figure}[tbp]
\includegraphics[scale=0.8]{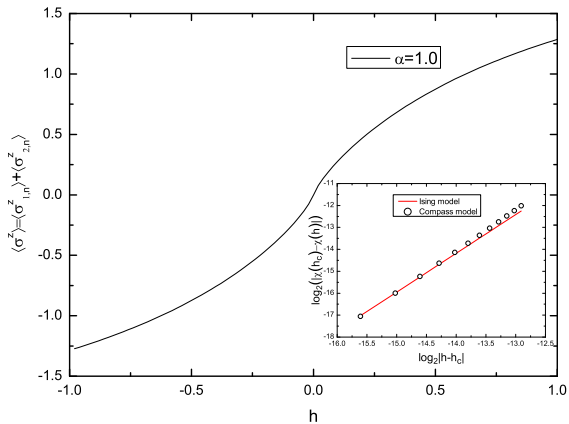}
\caption{ (Color online)  The pseudospin magnetization
($\langle\sigma^z\rangle=\langle\sigma_{1,n}^z\rangle+\langle\sigma_{2,n}^z\rangle$)
versus the transverse field for $\alpha=1.0$. Inset gives the
scaling of the value of $|\chi(h)-\chi(h_c)|$ as a function of
$|h-h_c|$. The red line denotes that in 1D transverse field Ising
model.} \label{order}
\end{figure}

\section{SUMMARY and DISCUSSION}

By using the method of mapping into a case with plural spin sites,
we obtain the exact GS energy and the GS wave function of 1D compass
model in a transverse magnetic field. The pseudo-spin liquid
disordered ground state is the universal features in the 1D compass
model. Meanwhile, we observe the second-order QPTs at
($\alpha=\pm1,h=0$). The energy gap $\Delta$ will survive even in
the thermodynamic limit for $h\ne0$. It is useful for supporting
naturally robust quantum information. The fidelity, the FS, the
concurrence, and the block-block entanglement entropy are also
calculated in terms of the obtained exact GS wave functions. The
finite-size scaling analysis suggests the second-order QPT occurs by
driving the transverse field. The pseudo-spin correlation functions,
the distance dependence of the pseudo-spin correlators and the
magnetization are also calculated. It is observed that the distance
dependence of $ \langle\sigma_{2,1}^x\sigma_{1,2+r}^x\rangle$
correlator displays a divergent correlation length when approaching
the critical points. The obtained scaling exponents are nearly the
same as those in the 1D transverse field Ising model, suggesting
that these two models share the same universality class. The scaling
exponent $c=1.0$ of the block-block entanglement entropy is the same
as the critical XX chain with no magnetic field, which shows the
different topological structure from the quantum Ising model. For
the 2D compass model with a transverse field, the degeneracy of GS
is removed because of the destruction of the symmetries. It is
expected that the QPT becomes more weak and the first-order QPT is
unlikely.

\section*{ACKNOWLEDGEMENTS}

We acknowledge useful discussions with Prof.  Lei-Han Tang. We also
thank Prof. Perk for pointing out one problem in the original
version of this paper. This work was supported by National Natural
Science Foundation of China, PCSIRT (Grant No. IRT0754) in
University in China, National Basic Research Program of China (Grant
No. 2009CB929104), Zhejiang Provincial Natural Science Foundation
under Grant No. Z7080203, and Program for Innovative Research Team
in Zhejiang Normal University.

$*$ Corresponding author. Email:qhchen@zju.edu.cn


\begin{references}
\bibitem{Dorier} J. Dorier, F. Becca, and F. Mila, Phys. Rev. B \textbf{72}, 024448
(2005).
\bibitem{HDChen} H. D. Chen, C. Fang, J. P. Hu, and H. Yao, Phys. Rev. B \textbf{75}, 144401
(2007).
\bibitem{Doucot} B.Doucot, M. V. Feigel'man, L. B. Ioffe, and A. S. Ioselevich, Phys. Rev. B \textbf{71},
024505 (2005).
\bibitem{Mishra} A. Mishra, M. Ma, F.-C. Zhang, S. Guertler, L.-H. Tang, S. L. Wan, Phys. Rev. Lett. \textbf{93}, 207201
(2004).
\bibitem{Brzezicki} W. Brzezicki, J. Dziarmaga, A. M.Ole$\acute{s}$, Phys. Rev. B \textbf{75}, 134415
(2007).
\bibitem{Ors} R. Or$\acute{u}$s, A. C. Doherty, and G. Vidal, Phys. Rev. Lett. \textbf{102},
077203 (2009).
\bibitem{Sun} K. W. Sun, Y. Y. Zhang and Q. H. Chen, Phys. Rev. B \textbf{79},
104429 (2009).
\bibitem{Eriksson}  E. Eriksson and H. Johannesson, Phys. Rev. B \textbf{79},
224424(2009).
\bibitem{You} W. L. You, and G. S. Tian, Phys. Rev. B \textbf{78},
184406 (2008).
\bibitem{Jackeli} G. Jackeli and G. Khaliullin, Phys. Rev. Lett. \textbf{102}, 017205 (2009).
\bibitem{Feng} X. Y. Feng, G. M. Zhang, and T. Xiang, Phys. Rev. Lett. \textbf{98},
087204 (2007).
\bibitem{Shi} X. F. Shi, Y. Yu, J. Q. You, and F. Nori, Phys. Rev. B \textbf{79},
134431 (2009).
\bibitem{Bombin} H. Bombin and M. A. Martin-Delgado, Phys. Rev. B \textbf{78},
115421 (2008).
\bibitem{Julien} J. Vidal, R. Thomale, K.P. Schmidt, and S. Dusuel, arXiv:0902.3547.
\bibitem{Scarola} V. W. Scarola, K. B. Whaley and M. Troyer, Phys. Rev. B \textbf{79},
085113 (2009).
\bibitem{Buonsante} P. Buonsante and A. Vezzani, Phys. Rev. Lett. \textbf{98}, 110601
(2007).
\bibitem{Cozzini} M. Cozzini, R. Ionicioiu, and P. Zanardi, Phys. Rev. B \textbf{ 76},
104420 (2007).
\bibitem{Chen} S. Chen, L. Wang, Y. J. Hao, and Y. P. Wang, Phys. Rev. A \textbf{ 77}, 032111
(2008).
\bibitem{Preskill} J. Preskill, J. Mod. Opt. \textbf{47}, 127 (2000).
\bibitem{Osborne} T. J. Osborne and M. A. Nielsen, Phys. Rev. A \textbf{66}, 032110
(2002); A. Osterloh et al., Nature (London) \textbf{416}, 608
(2002).
\bibitem{Vidal} G. Vidal \emph{et al.}, Phys. Rev. Lett. \textbf{90}, 227902 (2003); G.
Vidal, \emph{ibid}. \textbf{99}, 220405 (2007).
\bibitem{Korepin} V. E. Korepin, Phys. Rev. Lett. \textbf{92}, 096402 (2004); G. C.
Levine, \emph{ibid}. \textbf{93}, 266402 (2004); G. Refael and J.
E. Moore, \emph{ibid}. \textbf{93}, 260602 (2004); P. Calabrese
and J. Cardy, J. Stat. Mech. (2004) P06002.
\bibitem{Kitaev} A. Kitaev and J. Preskill, Phys. Rev. Lett. \textbf{96}, 110404
(2006); M. Levin and X.-G. Wen, \emph{ibid}. \textbf{96}, 110405
(2006).
\bibitem{Verstraete} F. Verstraete, M. A. Martin-Delgado, and J. I. Cirac,
Phys. Rev. Lett. \textbf{92}, 087201 (2004); W. Dur \emph{et al}.,
\emph{ibid}. \textbf{94}, 097203 (2005); H. Barnum \emph{et al}.,
\emph{ibid}. \textbf{92}, 107902 (2004).
\bibitem{Zhou}H. Q. Zhou, R. Or$\acute{u}$s, and G. Vidal, Phys. Rev. Lett. \textbf{100}, 080602
(2008).
\bibitem{Abasto} D. F. Abasto, A. Hamma, and P. Zanardi, Phys. Rev. A \textbf{77}, 022327 (2008).
\bibitem{Gu} S. J. Gu, H. M. Kwok, W. Q. Ning, and H. Q. Lin, Phys. Rev. B \textbf{77},
245109 (2008).
\bibitem{liu} T. Liu, Y. Y. Zhang, Q. H. Chen,   and K. L. Wang, Phys. Rev. A \textbf{80},
023810(2009)
\bibitem{Nielsen} M. Nielsen and I. Chuang, Quantum Computing and Quantum Communication
 (Cambridge Univ. Press, Cambridge, 2000).
\bibitem{Zhang} L. F. Zhang and P. Q. Tong, J. Phys. A \textbf{38}, 7377
(2005).
\bibitem{Osterloh}  A. Osterloh, L. Amico, G. Falci, and R. Fazio, Nature
(London) \textbf{416}, 608(2002); S. J. Gu, H. Q. Lin, and Y. Q. Li,
 Phys. Rev. A  \textbf{ 68}, 042330(2003).

\bibitem{Emary} N. Lambert, C. Emary,  and T. Brandes, Phys. Rev. Lett. \textbf{92},
073602(2004).
\bibitem{Liberti}  G. Liberti,  F.  Plastina, and F. Piperno, Phys. Rev.
A \textbf{74}, 022324 (2006).  J. Vidal and S. Dusuel, Europhys.
Lett. \textbf{74},  817(2006)).
\bibitem{chenqh} Q. H. Chen, Y. Y. Zhang, T.  Liu, and K. L. Wang,
Phys. Rev. A \textbf{78}, 051801(R) (2008).
\bibitem{Sasaki} S. Sasaki, Phys. Rev. E \textbf{53}, 168
(1996).
\bibitem{Perk} J. H. H. Perk, H. W. Capel, M. J. Zuilhof and Th. J. Siskens,
Physica A \textbf{81}, 319-348(1975).
\bibitem{Sachdev} S. Sachdev, \emph{Quantum Phase Transitions} (Cambridge University Press, Cambridge, England, 1999).
\bibitem{Lou} P. Lou and J. Y. Lee, Phys. Rev. B \textbf{74},
134402 (2006).
\bibitem{Refael} G. Refael and J. E. Moore, Phys. Rev. B. \textbf{76},
024419 (2007).
\bibitem{Fradkin} E. Fradkin and J. E. Moore, Phys. Rev. Lett. \textbf{97}, 050404
(2006).
\end{references}
\end{document}